\documentclass[12pt]{article}
\usepackage[dvips]{graphicx}
\usepackage{latexsym}
\parskip=\medskipamount

\newcommand {\cA}{{\cal A}}

\newcommand {\cD}{{\cal D}}

\newcommand {\cF}{{\cal F}}

\newcommand {\cL}{{\cal L}}
\newcommand {\cM}{{\cal M}}
\newcommand {\cN}{{\cal N}}

\newcommand {\cS}{{\cal S}}

\newcommand {\cW}{{\cal W}}

\newcommand{\bQ}{{\bf Q}}
\newcommand{\bR}{{\bf R}}

\def\a{\alpha}
\def \bi{\bibitem}

\def\b{\beta}

\def\d{\delta}

\def\G{\Gamma}

\def\l{\lambda}
\def\m{\mu}

\def\o{\omega}
\def\p{\pi}
\def\q{\theta}
\def\r{\rho}
\def\s{\sigma}
\def\t{\tau}

\def\z{\zeta}
\def\D{\Delta}
\def\F{\Phi}
\def\J{\Psi}
\def\L{\Lambda}
\def\O{\Omega}

\def\S{\Sigma}
\def\U{\Upsilon}

\def\tr{{\rm tr}}
\newcommand{\ad}{{\dot{\alpha}}}                           %new
\newcommand{\bd}{{\dot{\beta}}}                            %new
\newcommand{\ve}{\varepsilon}                            %new
\newcommand{\pa}{\partial}                           %new
\newcommand{\hf}{\frac12}
\newcommand{\vf}{\varphi}
\newcommand{\sect}[1]{\setcounter{equation}{0}\section{#1}}

\newcommand{\be}{\begin{equation}}
\newcommand{\ee}{\end{equation}}
\newcommand{\bea}{\begin{eqnarray}}
\newcommand{\eea}{\end{eqnarray}}
\newcommand{\non}{\nonumber}
\begin{document}
%%%%%%%%%%%%%%%%
%%%%%%%%%%%%%%%%

\begin{titlepage}
\thispagestyle{empty}

\begin{flushright}
hep-th/0308136 \\
August, 2003 \\
\end{flushright}
\vspace{5mm}

\begin{center}
{\Large \bf
Low-energy dynamics in N = 2 super  QED: \\
Two-loop approximation}
\end{center}
%\vspace{3mm}

\begin{center}
{\large S. M. Kuzenko and   I. N. McArthur}\\
\vspace{2mm}

\footnotesize{
{\it School of Physics, The University of Western Australia\\
Crawley, W.A. 6009, Australia}
} \\
{\tt  kuzenko@cyllene.uwa.edu.au},~
{\tt mcarthur@physics.uwa.edu.au} \\
\end{center}
\vspace{5mm}

\begin{abstract}
\baselineskip=14pt
The two-loop (Euler-Heisenberg-type) effective action 
for $\cN=2$ supersymmetric QED is computed
using the $\cN=1$ superspace formulation. 
The effective action is expressed as a series in
supersymmetric extensions of $F^{2n}$, where $n=2,3,\ldots$, 
with $F$ the  field strength. The corresponding coefficients 
are given by triple proper-time integrals which are evaluated 
exactly. As a by-product,  we demonstrate the appearance 
of a non-vanishing $F^4$ quantum correction at  the 
two-loop order. The latter result is in 
conflict with the conclusion of hep-th/9710142 
that no such quantum corrections 
are generated at two loops in generic $\cN=2$ SYM 
theories on the Coulomb branch. We explain a subtle loophole 
in the relevant consideration of hep-th/9710142 and re-derive 
the $F^4$ term  from harmonic supergraphs.
\end{abstract}

\vfill
\end{titlepage}

\newpage
\setcounter{page}{1}

\renewcommand{\thefootnote}{\arabic{footnote}}
\setcounter{footnote}{0}
%%%%%%%%%%%%%%%%%%%%%%%%%%%
\sect{Introduction and outlook}
In our recent paper \cite{KM}, 
a manifestly covariant approach was developed 
for evaluating multi-loop quantum corrections
to low-energy effective actions within the 
background field formulation. This approach 
is  applicable to ordinary gauge theories
and  to supersymmetric Yang-Mills theories
formulated in superspace. Its power is not restricted
to computing just the counterterms --  it is well suited
for deriving finite quantum corrections in 
the framework of the derivative 
expansion. More specifically, in the case of 
supersymmetric Yang-Mills theories, 
it is free of  some drawbacks 
still present  in the classic works \cite{GZ} 
(such as the splitting  
of  background covariant derivatives 
into ordinary derivatives plus the  background 
connection, in the process of evaluating 
the supergraphs).

As a simple application of the techniques 
developed in \cite{KM}, in this note
we derive  the two-loop 
(Euler-Heisenberg-type \cite{EH,W,Schwinger}) 
effective action for $\cN=2$ supersymmetric QED 
formulated in  $\cN=1$ superspace. 
This is a  supersymmetric generalization 
of the two-loop QED calculation by 
Ritus \cite{Ritus}  (see also follow-up publications
\cite{BS,DR,RSS,FRSS,KS}).
It is curious that  
the two-loop QED effective action 
\cite{Ritus} was computed  only a year 
after the work by Wess and Zumino 
\cite{WZ} that stimulated widespread interest in 
supersymmetric  quantum field theory.
To the best of our knowledge, 
the Ritus results have never been extended 
before to the supersymmetric case\footnote{At 
the component level, the two-loop effective action 
(\ref{two-loop-action}) is not just a combination 
of the Ritus  results for scalars and spinors \cite{Ritus}
because of the presence of   
quartic scalar and Yukawa 
couplings in $\cN=2$ SQED.}. 

Our interest in $\cN=2$ SQED, and not the 
`more realistic' $\cN=1$ SQED, is motivated by 
the fact that there exist  numerous 
(AdS/CFT-correspondence inspired) conjectures
about the multi-loop structure of 
(Coulomb-branch) low-energy 
actions in {\it extended} superconformal theories, 
especially the $\cN=4$ SYM theory, 
see, e.g. \cite{BPT} for a discussion and references.
None of these conjectures are  related to $\cN=2$ 
SQED which is, of course, not a superconformal theory. 
We believe, nevertheless, that the experience gained
and lessons learned
through the study of $\cN=2$ SQED should be an important
stepping stone towards testing these conjectures.

An unexpected outcome of the consideration in this 
paper concerns one particular conclusion drawn 
in \cite{BKO} on the basis of the background field 
formulation in $\cN=2$ harmonic superspace \cite{BBKO}.
According to \cite{BKO},  no $F^4$ quantum correction 
occurs at two loops in generic $\cN=2$ 
super Yang-Mills theories on the Coulomb branch, 
in particular in $\cN=2$ SQED. However, as 
it will be shown below,  on the basis of the background 
field formulation in $\cN=1$ superspace, 
there does occur a non-vanishing $F^4$ 
two-loop correction in $\cN=2$ SQED. 
Unfortunately, the analysis in  \cite{BKO} 
turns out to contain a subtle loophole related
to the intricate structure of harmonic supergraphs. 
A more careful treatment of two-loop harmonic supergraphs, 
which will be given in the present paper, 
leads to the same non-zero $F^4$  term 
 in $\cN=2$ SQED at two loops as that derived 
using the $\cN=1$ superfield formalism.
 
Some time ago, Dine and Seiberg \cite{DS} 
argued that the $F^4$ quantum correction is one-loop 
exact on the Coulomb branch of  $\cN=2,4$ 
superconformal theories.  It was also shown 
\cite{DKMSW,BFMT} that there are no  
instanton $F^4$ corrections. 
The paper \cite{BKO} provided perturbative two-loop 
support for the Dine-Seiberg conjecture. 
Since the two-loop $F^4$ conclusion of \cite{BKO}
is no longer valid, it would be extremely interesting 
to carry out  an 
independent calculation of the two-loop $F^4$ 
quantum correction 
in $\cN=2$ superconformal theories 
(it definitely vanishes in $\cN=4$ SYM).

 The present paper is organized as follows. 
In section 2  we review, following \cite{KM}, 
the structure of exact superpropagators in 
a covariantly constant $\cN=1$ vector 
multiplet background.  Section 3 
contains the  $\cN=2$ SQED setup
required for the subsequent consideration.
 The one-loop effective action for  $\cN=2$ SQED
is reviewed in section 4.
The two-loop effective action for  $\cN=2$ SQED
is derived in section 5 -- the main original part of 
this work.
In section 6 we re-derive the two-loop $F^4$ 
quantum correction using the harmonic superspace
 formulation for $\cN=2$ SQED.
The salient properties of the $\cN=1$ 
parallel displacement propagator 
are collected in appendix.

\sect{Exact superpropagators}
In this section we review, following \cite{KM}, 
the structure of exact superpropagators in a covariantly constant 
$\cN=1$ vector multiplet background. Our consideration 
is not restricted 
to the $U(1)$ case and is in fact valid for an arbitrary gauge group.
The results of this section can be used for loop calculations,
in the framework of the background field approach, 
of special sectors of low-energy effective actions in 
generic $\cN=1$ super Yang-Mills theories.
They will be used in the next sections to derive the two-loop 
(Euler-Heisenberg-type) effective action for $\cN=2$ SQED.

Green's functions in $\cN=1$ super Yang-Mills theories are 
typically associated with covariant  d'Alembertians constructed 
in terms of the
relevant  gauge  covariant derivatives
\be
\cD_A = (\cD_a, \cD_\a , {\bar \cD}^\ad ) 
= D_A +{\rm i}\cA_A(z)~,
\ee
with $D_A$ the flat covariant derivatives\footnote{Our $\cN=1$ notation 
and conventions correspond to \cite{BK}.}, and $\cA_A(z)$ 
the superfield connection taking its values in the Lie algebra of 
the gauge group.  
So we start by recalling the algebra of  gauge covariant 
derivatives:
\bea
& \{ \cD_\a , \cD_\b \} 
= \{ {\bar \cD}_\ad , {\bar \cD}_\bd \} =0~, \qquad 
\{ \cD_\a , {\bar \cD}_\bd \} = - 2{\rm i} \, \cD_{\a \bd}~, \non \\
& [ \cD_\a , \cD_{\b \bd}] = 2 {\rm i} \ve_{\a \b}\,{\bar \cW}_\bd ~, 
\qquad 
[{\bar \cD}_\ad , \cD_{\b \bd}] = 2{\rm i} \ve_{\ad \bd}\,\cW_\b ~ , 
\non \\
& [ \cD_{\a \ad}, \cD_{\b \bd} ] = {\rm i} \cF_{\a \ad, \b\bd} 
= - \ve_{\a \b}\, {\bar \cD}_\ad {\bar \cW}_\bd 
-\ve_{\ad \bd} \,\cD_\a \cW_\b~. 
\label{N=1cov-der-al}
\eea
Here  the spinor field strengths $\cW_\a$ and ${\bar \cW}_\ad$ 
obey the Bianchi  identities
\be
{\bar \cD}_\ad \cW_\a =0~, \qquad 
\cD^\a \cW_\a = {\bar \cD}_\ad {\bar \cW}^\ad~.
\ee

There are three major d'Alembertians which occur 
in covariant supergraphs
\cite{GGRS}: (i) the vector d'Alembertian ${\Box}_{\rm v} $; 
(ii) the chiral d'Alembertian $\Box_+$; and (iii) the antichiral 
d'Alembertian $\Box_-$.
The vector d'Alembertian is defined by
\bea 
{\Box}_{\rm v} 
&=& \cD^a \cD_a - \cW^\a \cD_\a +{\bar \cW}_\ad {\bar \cD}^\ad \\
&=& -\frac{1}{8} \cD^\a {\bar \cD}^2 \cD_\a 
+{1 \over 16} \{ \cD^2 , {\bar \cD}^2 \} 
-\cW^\a \cD_\a -\hf  (\cD^\a \cW_\a) \non \\
&=& 
 -\frac{1}{8} {\bar \cD}_\ad \cD^2 {\bar \cD}^\ad 
+{1 \over 16} \{ \cD^2 , {\bar \cD}^2 \} 
+{\bar \cW}_\ad {\bar \cD}^\ad +\hf({\bar \cD}_\ad {\bar \cW}^\ad ) ~.
\non 
\eea
Among its important properties are the identities
\be
\frac{1}{16} [\cD^2 , {\bar \cD}^2 ] = 
{\Box}_{\rm v} 
+\frac{\rm i}{2} {\bar \cD}_\ad \cD^{\a \ad } \cD_\a 
= - {\Box}_{\rm v} 
-\frac{\rm i}{2} \cD_\a \cD^{\a \ad } {\bar \cD}_\ad~. 
\label{iden1}
\ee
The  covariantly chiral d'Alembertian is defined by
\bea 
\Box_+ &=& \cD^a \cD_a - \cW^\a \cD_\a -\hf \, (\cD^\a \cW_\a)~, 
\quad
\Box_+ \F = {1 \over 16} \, {\bar \cD}^2 \cD^2 \F ~, \quad 
{\bar \cD}_\ad \F =0~.
\eea
As can be seen, the operator $\Box_+ $ acts on the space of 
covariantly chiral  superfields.
The antichiral d'Alembertian is defined similarly, 
\bea 
\Box_- &=& \cD^a \cD_a + {\bar \cW}_\ad {\bar \cD}^\ad 
+\hf \, ({\bar \cD}_\ad  {\bar \cW}^\ad)~, 
\quad
\Box_- {\bar \F} = {1 \over 16} \, \cD^2 {\bar \cD}^2  {\bar \F} ~, 
\quad  \cD_\a {\bar \F} =0~. 
\eea
The operators $\Box_+$ and $\Box_-$ are related to each other 
as follows:
\be
\cD^2 \,\Box_+ = \Box_- \, \cD^2~,  \qquad 
{\bar \cD}^2 \,\Box_- = \Box_+ \, {\bar \cD}^2~.
\ee
Additional relations occur for an on-shell background
\be
\cD^\a \cW_\a = 0 \quad \longrightarrow \quad 
\cD^2 \,\Box_+  = \cD^2 \,\Box_{\rm v} = \Box_{\rm v}\, \cD^2~, 
\quad
{\bar \cD}^2 \,\Box_-  = {\bar \cD}^2 \,\Box_{\rm v} 
= \Box_{\rm v} \, {\bar \cD}^2~.  
\ee
In what follows, the background vector multiplet 
is chosen to  be covariantly constant and on-shell,
\be
\cD_a \cW_\b = 0~, \qquad 
\cD^\a \cW_\a = 0~.
\ee
It is worth noting that the first requirement here implies that 
the Yang-Mills superfield belongs to  the Cartan subalgebra
of the gauge group.

Associated with $\Box_{\rm v} $ is a Green's function $G(z,z')$ 
which is subject to the Feynman boundary conditions 
and satisfies the equation 
\be
\Big(\Box_{\rm v} - m^2 \Big) \, G(z,z') = - {\bf 1}\,\d^8 (z-z')~.
\label{real-green}
\ee
It possesses the proper-time representation 
\be
G(z,z') = {\rm i} \int\limits_0^\infty {\rm d}s \, K(z,z'|s) \, 
{\rm e}^{ -{\rm i} (m^2 -{\rm i}\ve ) s }~, 
\qquad   \ve \to +0~.
\label{proper-time-repr}
\ee
The corresponding heat kernel\footnote{This
heat kernel was first derived  in the Fock-Schwinger 
gauge in \cite{O}.}
\cite{KM} is  
\bea
K(z,z'|s) &=& -\frac{\rm i}{(4 \pi s)^2} \, 
\sqrt{
\det
\left( \frac{2\, s \,\cF}{{\rm e}^{ 2  s \cF} -1}\right) } 
\; {\rm U}(s) \,
\z^2  \bar{\z}^2 \,
{\rm e}^{ \frac{{\rm i}}{4} 
\r \, \cF \coth ( s \cF) \, \r } \, I(z,z')~,
\label{real-kernel}
\eea
where the determinant is computed
with respect to the Lorentz indices,
\be 
{\rm U}(s) = \exp \Big\{- {\rm i} s (\cW^{\a} \cD_{\a} 
+ \bar{\cW}^{\ad} {\bar \cD}_{\ad})\Big\}~,
\ee
and $I(z,z') $ is the so-called parallel displacement propagator, 
see the Appendix for its definition and basic properties. 
The supersymmetric two-point function 
$\z^A(z,z') =- \z^A(z',z)=(\r^a , \z^\a, {\bar \z}_\ad)$
is defined as follows: 
\be
\r^a = (x-x')^a - {\rm i} (\q-\q') \s^a {\bar \q}' 
+ {\rm i} \q' \s^a ( {\bar \q} - {\bar \q}') ~, \quad
\z^\a = (\q - \q')^\a ~, \quad
{\bar \z}_\ad =({\bar \q} -{\bar \q}' )_\ad ~. 
\label{two-point}
\ee
Let us introduce proper-time dependent variables
$\J(s) \equiv {\rm U}(s) \, \J \, {\rm U}(-s)$. 
With the notation 
\be
\cN_\a{}^\b = \cD_\a \cW^\b~, \qquad 
{\bar \cN}_\ad{}^\bd = {\bar \cD}_\ad {\bar \cW}^\bd~, 
\qquad {\rm tr} \, \cN = {\rm tr} \, {\bar \cN} =0~,
\ee
for the buiding blocks appearing in the right hand side of 
(\ref{real-kernel}) we then get
\bea
 \cW^{\a}(s) 
= (\cW\, {\rm e}^{- {\rm i} s \cN} )^{\a}~, 
\quad && \quad
{\bar \cW}^{\ad}(s) 
= ({\bar \cW}\, {\rm e}^{- {\rm i} s {\bar \cN}} )^{\ad}~, 
\nonumber \\
 \z^{\a}(s) 
= \z^{\a} + \Big( \cW \,
\frac{  {\rm e}^{ -{\rm i} s \cN} -1} {\cN} \Big)^{\a}~,
\quad && \quad 
{\bar \z}^{\ad}(s) 
= {\bar \z}^{\ad} - \Big( {\bar \cW} \,
\frac{  {\rm e}^{-{\rm i}s {\bar \cN}} -1} {\bar \cN}\Big)^{\ad}~, 
\label{zeta(s)}  \\
\r_{\a \ad}(s) 
= \r_{\a \ad} &-&2 \int_0^{s} {\rm d}t \, \Big( \cW_{\a}(t) \bar{\z}_{\ad}(t) 
+ \z_{\a}(t)\bar{\cW}_{\ad}(t) \Big)~.
\non
\eea
One also finds \cite{KM}
\be
U(s) \, I(z,z') = \exp \,\Big\{
\int_0^s {\rm d}t \, \Xi(\z(t), \cW(t),
\bar{\cW}(t)) \Big\} \, I(z,z')~,
\ee
where $\Xi(\z(s), \cW(s),  \bar{\cW}(s)) = U(s) \, \Xi(\z, \cW,
\bar{\cW}) \, U(-s)$ and 
\bea
\Xi (\z, \cW,
\bar{\cW}) &=& \frac{1}{12} \, \r^{\ad \a}\Big( \cW^{\b}\bar{\z}^{\bd} 
- \z^{\b} \bar{\cW}^{\bd}\Big) \Big(\ve_{\b \a} \, 
\bar{\cD}_{\bd} {\bar \cW}_{\ad} 
- \ve_{\bd \ad} \, \cD_{\b}\cW_{\a}\Big) 
-\frac{2 {\rm i}}{3}\, \z \cW  \, \bar{\z} \bar{\cW} 
\nonumber \\
&-& {{\rm i} \over 3} \, \z^2 \, \Big( \bar{\cW}^2
- \frac{1}{4} \bar{\z}
\bar{\cD} \,\bar{\cW}^2 
\Big) 
- {{\rm i} \over 3}\, \bar{\z}^2 \, \Big(  \cW^2 - \frac{1}{4} \z
\cD \,\cW^2  \Big)~.
\eea

In the case of a real representation of the gauge group, 
the Green's function $G(z,z')$ should be realizable as 
the vacuum average of  a time-ordered product,
$$G(z,z') = {\rm i}\,\langle 0| T \, \Big(\S(z)\,  \S^{\rm T} (z') \Big) |0\rangle
\equiv  {\rm i}\,\langle \S(z) \, \S^{\rm T} (z')\rangle~,$$ 
for a real quantum field $\S(z)$. Therefore the corresponding heat 
kernel should possess the property 
\be 
K(z', z|s) = K^{\rm T}(z,z'|s)~.
\label{transpose}
\ee
As is seen from (\ref{real-kernel}), this property
is only obvious for the sub-kernel $\tilde{K}(z,z'|s) $ defined by 
\be
K(z,z'|s) = {\rm U}(s) \, \tilde{K}(z,z'|s) ~.
\ee
However, using the properties of the parallel displacement 
propagator listed in the Appendix, 
one can show
\be
\Big( \cW^\a \cD_\a + {\bar \cW}^\ad {\bar \cD}_\ad \Big)
\tilde{K}(z,z'|s) = \tilde{K}(z,z'|s) 
\Big( \stackrel{\leftarrow}{\cD'_\a} \cW'^\a
+ \stackrel{\leftarrow}{{\bar \cD}'_\ad} {\bar \cW}'^\ad \Big)~,
\ee
and this in fact implies (\ref{transpose}).

Associated with the  chiral d'Alembertian $\Box_+$ 
is a Green's function $G_+(z,z'|s)$ which is covariantly 
chiral in both arguments, 
\be
{\bar \cD}_\ad \, G_+(z,z') = {\bar \cD}'_\ad \, G_+(z,z') =0~,
\ee
is subject to the Feynman boundary conditions and satisfies
the equation 
\be
\Big(\Box_+ - m^2 \Big) \, G_+(z,z') = - {\bf 1}\,\d_+ (z,z')~, 
\qquad {\bf 1}\,\d_+ (z,z') = -{1\over 4}{\bar \cD}^2\, {\bf 1}\,\d^8(z-z')~.
\ee
Under the restriction $\cD^\a \cW_\a=0$, this Green's function 
 is related to $G(z,z')$ as follows:
\be 
G_+(z,z') = -{1 \over 4} {\bar \cD}^2 G(z,z')
= -{1 \over 4} {\bar \cD}'^2 G(z,z')~.
\ee
The corresponding chiral heat kernel\footnote{In 
the $U(1)$ case, the chiral heat kernel 
was first derived in a special gauge in \cite{BK2}.}
 turns out to be 
\bea
K_+(z,z'|s) &=& -{1 \over 4} {\bar \cD}^2 K(z,z'|s) =
-\frac{\rm i}{(4 \pi s)^2} \, 
\sqrt{ \det
\left( \frac{2\, s \,\cF}{{\rm e}^{ 2  s \cF} -1}\right) } 
\; {\rm U}(s) \,
\non \\
& \times &
\z^2   \,
\exp 
\Big( 
 \frac{{\rm i}}{4} \r \,  
\cF \coth ( s \cF) \, \r
-\frac{{\rm i}}{2} \r^a 
\cW   \s_a {\bar \z} \Big)
 \,  I(z,z')~.
\label{chiral-kernel}
\eea
It is an instructive exercise to check, using the properties
of the parallel displacement propagator given in the Appendix, 
that $K_+(z,z'|s) $ is covariantly chiral in both arguments.
${}$For completeness, we also present the antichiral-chiral 
kernel 
\bea
\frac{1}{16} \, \cD^2  \bar{\cD}'^2\,  K(z,z'|s)
& = &  - \, \frac{ {\rm i} }{(4 \pi s)^2} \,
\sqrt{
\det \left( \frac{2 s \cF}{{\rm e}^{2s \cF}-1}
\right)
}
\; {\rm U}(s)  \non \\ 
&  \times &  
\exp \Big( \frac{ \rm  i }{4} \tilde{\r}
  \, \cF \coth (s \cF)
\, \tilde{\r} +R (z,z') \Big)  \, I(z,z')~ ,
\eea
where
\bea
R ( z,z' )  &=&  - \frac{{\rm i}}{2} \, \tilde{\r}^a
( \cW \s_a 
\bar{\z}  + \z \s_a 
\bar{\cW})  
+ \frac13 (
 \z^2 \,\bar{\z} \bar{\cW} -  \bar{\z}^2 \, \z \cW ) \non \\
&&+ \frac{{\rm i}}{12} \tilde{\r}_{\a \ad} ( \z^{\a} 
\bar{\z}^{\bd} \, \bar{\cD}_{\bd} \bar{\cW}^{\ad}
+ 5\,   \z^{\b}  \bar{\z}^{\ad}\,
\cD_{\b} \cW^{\a}) ~,
\eea
and $\tilde{\r}^a$ is a `left antichiral/right chiral'  variable
\be
\tilde{\r}^a = \r^a -  {\rm i} \, \z \s^a \bar{\z}~, \qquad
\cD_{\b} \, \tilde{\r}^a  = \bar{\cD}'_{\bd} \,
\tilde{\r}^a =0~ .
\ee

The parallel displacement propagator is the only 
building block for the supersymmetric heat kernels
which involves the naked gauge connection.
In covariant supergraphs, however, 
the parallel displacement propagators, 
that come from all possible internal lines, 
`annihilate' each other through the mechanism 
sketched in \cite{KM}.

A very special and extremely simple type of background 
field configuration,
\be
\cD_\a \cW_\b = 0~,
\label{glueball}
\ee
is suitable for computing exotic low-energy effective actions 
of the form
\be
\int{\rm d}^8z \, L(\cW, {\bar \cW}) ~+~
\Big( \int{\rm d}^6z \, P(\cW) ~+~ {\rm c.c.} \Big)~,
 \ee
which are of some interest in the context of  
the Veneziano-Yankielowicz action \cite{VY} and its 
recent generalizations destined to describe 
the low-energy dynamics
of the glueball superfield $\cS={\rm tr} \, \cW^2$. 
Under the constraint (\ref{glueball}),
the kernel (\ref{real-kernel}) becomes
\be
K(z,z'|s) = -\frac{\rm i}{(4 \pi s)^2} \, 
{\rm e}^{ {\rm i}\r^2 /4s } \, 
\d^2(\z  -{\rm i} s \,\cW) \,
\d^2({\bar \z}  + {\rm  i}s \,{\bar \cW} ) 
\, I(z,z')~,  
\ee
while the chiral kernel (\ref{chiral-kernel})
turns into\footnote{A simplified version of the chiral kernel
(\ref{chiral-glueball}) has recently  been used 
in \cite{DGLVZ} to provide further support
to the Dijkgraaf-Vafa conjecture \cite{DV}.}
\be
K_+(z,z'|s) = -\frac{\rm i}{(4 \pi s)^2} \, 
{\rm e}^{ {\rm i}\r^2 /4s } \, 
\d^2(\z  -{\rm i} s \,\cW) \,
{\rm e}^{ \frac{{\rm i}}{6} s\,\cW^2 \,({\bar \z} +{\rm i}s\, {\bar \cW})^2}
%{ -\frac{{\rm i}}{2} \r^a 
%\cW   \s_a ({\bar \z} +{\rm i}s\, {\bar \cW})}
\, I(z,z')~. 
\label{chiral-glueball}
\ee
 Here the parallel displacement propagator is completely 
specified by the properties:
\bea
I(z',z) \, \cD_{\a \ad} I(z,z') &=&
-{\rm i} ( \z_{\a} \bar{\cW}_{\ad} 
+ \cW_\a \, \bar{\z}_{\ad}   ) ~, \non \\
I(z',z) \,  \cD_{\a} I(z,z') &=&
- \frac{{\rm i}}{2} \, \r_{\a \ad} 
\bar{\cW}^{\ad} 
+ \frac13 ( \z_{\a} \bar{\z} \bar{\cW}
+ \bar{\z}^2  \cW_{\a})~, \\
I(z',z) \,  {\bar \cD}_{\ad} I(z,z') &=&
- \frac{{\rm i}}{2} \, \r_{\a \ad} \cW^{\a} 
- \frac13 ( {\bar \z}_\ad \z \cW  + \z^2 {\bar \cW}_\ad)~. \non
\eea

\sect{\mbox{\bf $\cN = 2$} SQED}

The action of $\cN=2$ SQED written in terms of $\cN=1$ 
superfields is 
\bea
S_{\rm SQED} &=& \frac{1}{e^2} \int {\rm d}^8 z \, {\bar \F} \F
+  \frac{1}{e^2} \int {\rm d}^6 z \, W^\a W_\a \non \\
&& +\int {\rm d}^8 z \, \Big( \overline{Q} {\rm e}^V Q 
+ \overline{ \tilde{Q} } {\rm e}^{-V} \tilde{Q} \Big)
+ \Big(
{\rm i}  \int {\rm d}^6 z \, \tilde{Q} \F Q + {\rm c.c.} \Big)~,
\label{n=2sqed-action0}
\eea
where $W_\a = - {1\over 8} {\bar D}^2 D_\a V$. 
The dynamical variables $\F$ and $V$ describe an
$\cN=2$ Abelian vector multiplet, while 
the superfields $Q$ and  $\tilde{Q}$ constitute  
a massless Fayet-Sohnius hypermultiplet.
The case of a massive hypermultiplet is obtained 
from (\ref {n=2sqed-action0}) by the shift
$\F \to \F+ m$, with $m$ a complex parameter.\footnote{The 
action of $\cN=1$ SQED is obtained from (\ref {n=2sqed-action0})
by discarding $\F$ as a dynamical variable, and instead
`freezing' $\F$ to a constant value $m$.}
Introducing new  chiral variables
\bea
\bQ= \exp \Big( {{\rm i} {\p \over 4} \s_1} \Big)
\left(
\begin{array}{c}
Q \\
\tilde{Q}  
\end{array}
\right)~,
\eea
with $\vec{\s} = (\s_1, \s_2, \s_3)$ the Pauli matrices, 
the action takes the (real representation) form
\bea
S_{\rm SQED} &=& \frac{1}{e^2} \int {\rm d}^8 z \, \bar{\F} \F
+  \frac{1}{e^2} \int {\rm d}^6 z \, W^\a W_\a \non \\
&& +\int {\rm d}^8 z \,  \bQ^\dagger  {\rm e}^{V\s_2}  \bQ 
+ \hf \Big(  \int {\rm d}^6 z \,  \F \, \bQ^{\rm T} \bQ  + {\rm c.c.} \Big)~.
\label{n=2sqed-action}
\eea

We are interested in a low-energy effective action
 $\G[W,  \F]$ which describes the dynamics of the $\cN=2$
massless vector multiplet and which is generated  by integrating 
out  the massive charged hypermultiplet. 
More specifically, we concentrate on a slowly 
varying part of $\G[W,  \F]$  that, at the component level, 
comprises contributions with (the supersymmetrization of) all 
possible powers of the gauge field strength without derivatives.
Its generic form is \cite{BKT}
\be
\G[W,  \F] = \Big(
\a \int{\rm d}^6 z \,W^2\, \ln {\F \over \m} 
~+ ~ {\rm c.c.} \Big) 
~+~ \int{\rm d}^8 z \,{ {\bar W}^2 W^2 \over  {\bar \F}^2\F^2 } 
\, \O(\J^2, {\bar \J}^2)~,
\label{structure}
\ee
where 
\be
{\bar \J}^2 = {1 \over4} D^2 \Big( { W^2  \over  {\bar \F}^2\F^2 } \Big)~,
\qquad 
\J^2 = {1 \over4} {\bar D}^2 \Big( { {\bar W}^2  \over  {\bar \F}^2\F^2 } 
\Big)~,
\label{psi}
\ee
$\m$ is the renormalization scale and 
$\O$ some real analytic function.
The first term on the right hand side of (\ref{structure})
is known to be one-loop exact in perturbation theory, 
while the second term receives quantum corrections 
at all  loops.

To evaluate quantum loop corrections 
to the effective action  (\ref{structure}), 
we use the $\cN=1$ superfield background field method
in its simplest realization, as we are dealing with an Abelian 
gauge theory.  Let us split the dynamical variables as follows:
$\F \to \F + \vf, ~ V \to V + v, ~ \bQ \to \bQ + {\bf q}$, where 
$\F, V$ and $\bQ$ are background superfields, 
while $\vf, v$ and $\bf q$
are quantum ones. As is standard in the background field approach, 
(background covariant) gauge conditions are to be introduced 
for the quantum gauge freedom while keeping intact
the background gauge invariance.  Since we are only interested 
in the quantum corrections of the form (\ref{structure}), 
it is sufficient to consider simple background configurations
\be
\pa_a W_\b = D^\a W_\a = 0~, \qquad 
D_\a \F = 0~, \qquad \bQ = 0~.
\ee
Upon quantization in Feynman gauge, 
we end up with the following action to be used for 
loop calculations 
\bea
S_{\rm quantum} &=& \frac{1}{e^2} \int {\rm d}^8 z \,
\Big( \bar{\vf} \vf - \hf \,v \Box v \Big)
\non \\
&& 
+\int {\rm d}^8 z \,  {\bf q}^\dagger {\rm e}^{v\,\s_2}  {\bf q} 
+ \hf \Big(  \int {\rm d}^6 z \,  (\F + \vf) \, 
{\bf q}^{\rm T} {\bf q}  + {\rm c.c.} \Big)~,
\label{quantum-action}
\eea
with $\Box = \pa^a \pa_a$.
It is  understood here that the quantum superfields $\bf q$
and ${\bf q}^\dagger$ 
are background covariantly chiral and antichiral, respectively, 
\be
{\bar \cD}_\ad {\bf q} = 0~, \qquad 
\cD_\a {\bf q}^\dagger =0~.
\ee

${}$From the quadratic part of  (\ref{quantum-action})
one reads off the Feynman propagators
\bea
{\rm i} \, \langle  {\bf q} (z) {\bf q}^\dagger (z') \rangle &=& 
{1 \over 16} {\bar \cD}^2 \cD'^2 \, G (z,z') ~, \non \\
{\rm i} \, \langle  {\bf q} (z)  {\bf q}^{\rm T} (z') \rangle &=& 
{1 \over 4} {\bar \F} \, 
 {\bar \cD}^2  G (z,z') 
= {1 \over 4} {\bar \F} \,  {\bar \cD}'^2  G (z,z') ~,
\label{propagators} \\
{\rm i} \, \langle  \vf (z) {\bar \vf} (z') \rangle &=& 
-{e^2 \over 16} {\bar D}^2 D'^2 \, \frac{1}{\Box} \,
\d^8 (z-z') 
= -{{\rm i} \over 16} {\bar D}^2 D'^2
\,   \langle  v (z) v (z') \rangle
~, \non \\
 {\rm i} \, \langle  v (z) v (z') \rangle &=& 
\frac{e^2}{\Box}\,  \d^8 (z-z')
= -{ e^2  \over (4\p)^2} \int_0^\infty  \frac{{\rm d}u}{u^2} \,
{\rm e}^{ {\rm i}\r^2 /4u } \, 
\d^2(\z ) \,\d^2( \bar \z ) ~. \non 
\eea
Here the Green's function $G(z,z') $ 
transforms in the defining representation of $SO(2) \cong U(1)$, 
and satisfies the equation (\ref{real-green}) with $m^2={\bar \F}\F$.
It is given by the proper-time representation 
(\ref{proper-time-repr})
with the heat kernel $K(z,z'|s)$  specified in (\ref{real-kernel}). 
It is understood that the field strengths $\cW_\a$, ${\bar \cW}_\ad$ 
and their covariant derivatives  (such as $\cF_{ab}$)
are related to $W_\a$, ${\bar W}_\ad$
as follows 
\be
\cW_\a = W_\a \, \s_2~, \qquad 
{\bar \cW}_\ad = {\bar W}_\ad \,\s_2~.
\ee

\sect{One-loop effective action}

${}$For the sake of completeness, we discuss here the structure of the 
one-loop effective action \cite{SY,O,MG,PB,BKT}. 
Its formal representation is (see \cite{BK} for more details)
\be
\G_{\rm one-loop} = -\frac{ \rm i }{2} \m^{2 \o} 
\int\limits_0^\infty  \frac{ {\rm d} ({\rm i} s) }{ ({\rm i} s)^{1-\o} } \,
{\rm Tr} \,K_+(s) \, {\rm e}^{ -{\rm i} ({\bar \F}\F -{\rm i}\ve ) s }~, 
\ee
where $\o$ is the regularization parameter
($\o \to 0$ at the end of calculation),
 and $\m$ the normalization 
point. The functional trace of the chiral kernel is defined by
\be 
{\rm Tr}\, K_+(s) = \int{\rm d}^6z\, {\rm tr} \, K_+(z,z|s)~.
\ee

Using the explicit form of the chiral kernel  
(\ref{chiral-kernel}),  we obtain 
\bea
\hf {\rm tr}\, K_+(z,z|s) = 
\frac{\rm i}{(4\p)^2 }\,W^2\,
\frac{\sin^2 (sB/2)}{(sB/2)^2} \,
\sqrt{ \det
\left( \frac{s F}{\sinh(s F) } \right) } ~,
\eea
where we have introduced the notation 
\bea
B^2 = \hf {\rm tr}\, N^2~, \quad N_\a{}^\b = D_\a W^\b~; \qquad
{\bar B}^2 = \hf {\rm tr}\, {\bar N}^2~, \quad 
{\bar N}_\ad{}^\bd = {\bar D}_\ad {\bar W}^\bd~.
\eea
${}$For the background superfields under consideration, 
we have 
\be
B^2 = {1 \over 4} D^2 W^2~, \qquad 
{\bar B}^2 = {1 \over 4} {\bar D}^2 {\bar W}^2~.
\label{b-barb}
\ee
The latter objects turn out to appear as building blocks for 
the eigenvalues of $F=(F_a{}^b)$ which are equal to $\pm \l_+$
and $\pm \l_-$, where 
\be
\l_\pm = \frac{\rm i}{2} (B \pm {\bar B})~, \qquad 
2 B^2 =  F^{ab}F_{ab} 
+ \frac{\rm i}{2} \ve^{abcd} F_{ab} F_{cd}~.
\label{eigenvalues}
\ee
This gives
\be
\sqrt{ \det
\left( \frac{s F}{\sinh(s F) } \right) } 
=\frac{ s\l_+}{\sinh (s\l_+)} \, \frac{ s\l_-}{\sinh (s\l_-)} ~.
\label{det1}
\ee
Now, the effective action takes the form 
\be
\G_{\rm one-loop} = 
\frac{ \m^{2 \o} }{(4\p)^2}
\int\limits_0^\infty  \frac{ {\rm d} ({\rm i} s) }{ ({\rm i} s)^{1-\o} } \,
 \int{\rm d}^6z\, W^2 \, \U(sB,s{\bar B})
 \, {\rm e}^{ -{\rm i} ({\bar \F}\F -{\rm i}\ve ) s }~,
\ee
where 
\be
\U(x,y) = \frac{ \cos x - 1}{x^2 }\, 
\frac{x^2-y^2}{\cos x - \cos y} ~, \qquad 
\U(x,0) = 0~.
\ee
Introducing a new function $\z(x,y)$ related to $\U$ by  \cite{BKT}
\be
\U(x,y) -1 = - y^2 \, \z(x,y)~, \quad 
\z(x,y)=\z(y,x) = 
\frac{ y^2(\cos x - 1) - x^2(\cos y -1)} {x^2 y^2(\cos x - \cos y)}
\ee
allows one to  readily  separate a UV divergent contribution 
and to  represent 
the finite part of the effective action as 
an integral over the full superspace.
Making use of eq. (\ref{b-barb}) and the standard identity
$$
- {1\over 4} \int {\rm d}^6 z \, {\bar D}^2\, L =  
\int {\rm d}^8 z \, L~, 
$$
for the renormalized one-loop effective action\footnote{In 
deriving the effective action (\ref{one-loop}), we concentrated 
on the quantum corrections
involving the $\cN=1$ vector multiplet field strength 
and did not take 
into account the effective K\"ahler potential 
$K({\bar \F},\F)= - \frac{1 }{(4\p)^2}
{\bar \F}\F \ln ({\bar \F} \F / \m^2)= 
{\bar \F} \cF'(\F) + \F{\bar \cF}'(\bar \F )$
generated by the holomorphic Seiberg potential 
$ \cF (\F) = - \frac{1 }{(4\p)^2} \F^2 \ln ( \F / \m)$, 
as well as higher derivative quantum corrections with  
chiral superfields. 
A derivation of $K({\bar \F},\F)$  using the superfield 
proper-time
technique was first given in \cite{BKY,BK}, see also more recent
calculations  \cite{dGR,PW} based on conventional supergraph 
techniques. The leading higher derivative quantum correction
with chiral superfields
was computed in \cite{BKY}.}
one ends up with 
\bea
\G_{\rm one-loop} &=& - \frac{1 }{(4\p)^2}
 \int{\rm d}^6z\, W^2 \, \ln {\F \over \m} ~+ ~{\rm c.c.} \non \\
&&+ \frac{1 }{(4\p)^2}
 \int{\rm d}^8z\, \frac{ {\bar W}^2 W^2}{{\bar \F}^2 \F^2} \,
\int\limits_0^\infty   {\rm d}  s \, s\,
\z(s \J,s{\bar \J}) \, {\rm e}^{ -{\rm i} (1 -{\rm i}\ve ) s }~,
\label{one-loop}
\eea
with $\J$ and $\bar \J$ defined in (\ref{psi}).

\sect{Two-loop effective action}
We now turn to computing the two-loop quantum correction to 
the effective action.  There are three supergraphs contributing 
at  two loops\footnote{There is actually one more 
two-loop supergraph, 
the so-called `eight' diagram, generated 
by the quartic vertex
$\hf \int{\rm d}^8z\, {\bf q}^\dagger v^2  {\bf q} $;
its contribution is obviously zero.}, 
and they are depicted in Figures 1--3.
\begin{figure}[!htb]
\begin{center}
\includegraphics{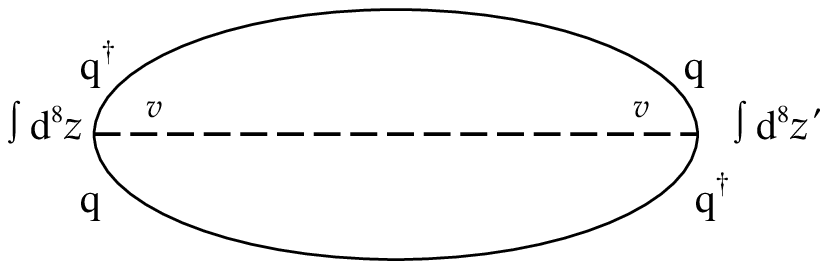}
\caption{Two-loop supergraph I}
\end{center}
\end{figure}
\begin{figure}[!htb]
\begin{center}
\includegraphics{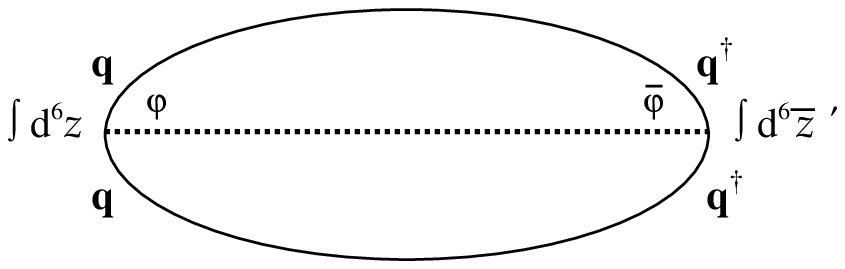}
\caption{Two-loop supergraph II}
\end{center}
\end{figure}
\begin{figure}[!htb]
\begin{center}
\includegraphics{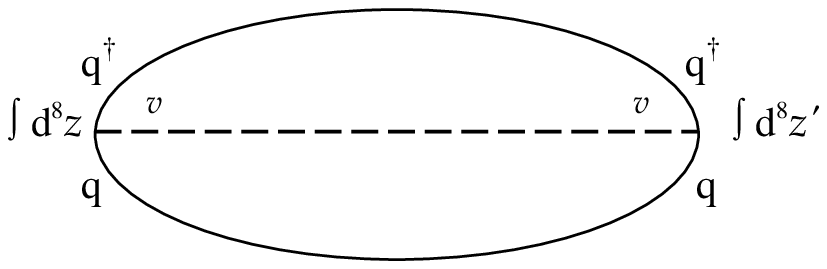}
\caption{Two-loop supergraph III}
\end{center}
\end{figure}

The  contribution from the first two supergraphs is  
\bea
\G_{\rm I +II} = {{\rm i} \over 2^9 } \int {\rm d}^8 z 
\int {\rm d}^8 z' \,  \langle  v (z) v (z') \rangle  
\, {\rm  tr} \left\{ 
\Big( {\bar \cD}^2 \cD^2 \, G(z,z') \Big) \,
 [{\bar \cD}^2, \cD^2 ] \,G(z',z) \right\}~.
\label{I+II}
\eea
The third supergraph leads to the following contribution
\bea
\G_{\rm III} = {{\rm i} \over 2^5 } \int {\rm d}^8 z 
\int {\rm d}^8 z' \,  \langle  v (z) v (z') \rangle  
\, {\bar \F}\F\, {\rm  tr} \left\{ 
\Big( {\bar \cD}^2  \, G(z,z') \Big) \,
  \cD^2  \,G(z',z) \right\}~.
\label{III}
\eea

It turns out that the expression for $\G_{\rm I +II}$ can be 
considerably simplified
using the properties of the  superpropagators and 
their heat kernels,  which were discussed in sect. 2. 
Since 
$$ 
{\bar \cD}^2 \, G(z,z') = {\bar \cD}'^2 \, G(z,z') ~, \qquad 
\cD^2 \, G(z,z') = \cD'^2 \, G(z,z')~, 
$$
we have 
$$
 {\bar \cD}^2 \cD^2 \, G(z,z') =  \cD'^2 {\bar \cD}'^2  \, G(z,z') ~, 
$$
and therefore 
\be
[{\bar \cD}^2, \cD^2 ] \,G(z,z') 
= - [{\bar \cD}'^2, \cD'^2 ] \, G(z,z') ~.
\ee
The latter relation in conjunction with the symmetry property
\be
 \langle  v (z) v (z') \rangle 
=  \langle  v (z') v (z) \rangle ~, \qquad 
G(z,z') = \Big( G(z',z) \Big)^{\rm T} 
\ee
leads to the new representation for $\G_{\rm I +II} $
\bea
\G_{\rm I +II} 
= {{\rm i} \over 2^{10} } \int {\rm d}^8 z 
\int {\rm d}^8 z' \,  \langle  v (z) v (z') \rangle  
\, {\rm  tr} \left\{ 
\Big( [{\bar \cD}^2, \cD^2 ]\, G(z,z') \Big) \,
 [{\bar \cD}^2, \cD^2 ] \,G(z',z) \right\}~.
\label{final1}
\eea
In accordance with (\ref{iden1}), we can represent 
\be
\frac{1}{16} [\cD^2 , {\bar \cD}^2 ] =   
\frac{\rm i}{4} {\bar \cD}_\ad \cD^{\a \ad } \cD_\a 
-\frac{\rm i}{4} \cD_\a \cD^{\a \ad } {\bar \cD}_\ad~,
\ee
and this identity turns out  to be very useful  
when computing the action 
of the commutators of covariant derivatives in (\ref{final1})
on  the Green's functions. A direct evaluation gives
\bea
\frac{1}{16} [ {\bar \cD}^2, \cD^2 ] \, K(z,z' | s) 
& \approx  &
\frac{\rm i}{(4 \pi s)^2} \, 
\sqrt{ \det
\left( \frac{2\, s \,\cF}{{\rm e}^{ 2  s \cF} -1}\right) } \,
\Big( \r \, \frac{2\cF}{{\rm e}^{ 2  s \cF} -1} \Big)^{\a \ad} \,
\z_\a (s)   \bar{\z}_\ad (s)  
\non \\
& \times &
{\rm e}^{ \frac{{\rm i}}{4} 
\r \, \cF \coth ( s \cF) \, \r } \, I(z,z') ~,
\eea
where we have omitted all terms of at least third order 
in the Grassmann variables  $\z_\a, \,{\bar \z}_\ad$ and 
$\cW_\a, \, {\bar \cW}_\ad$ 
as they  do not contribute to (\ref{final1}).
It is easy to derive 
\bea
\Big( \r \, \frac{2\cF}{{\rm e}^{ 2  s \cF} -1} \Big)^{\a \ad}\,
\z_\a (s)   \bar{\z}_\ad (s) 
 \Big|_{\z = {\bar \z}=0}
= \r \, \frac{F}{ \sinh( s F)} \, 
\frac{\sinh(s F_+)}{F_+} \, \frac{\sinh(s F_-)}{F_-}\J\, {\bf 1}_2~, 
\eea
where 
\be
\J^a = W \s^a {\bar W}~, \qquad
F_\pm = \hf ( F \pm {\rm i}\, \tilde{F})~,
\ee
with $\tilde{F}$ the Hodge-dual of $F$.
Here we have taken into account the fact  that 
$\cF = F\, \s_2$. 

As the propagator $ \langle  v (z) v (z') \rangle  $
contains the Grassmann delta-function 
$\d^2(\z)  \d^2(\bar \z )$, 
the integral over $\q'$ in (\ref{final1}) can be trivially done. 
Replacing  the bosonic integration variables 
in (\ref{final1}) by the rule $\{x, \, x' \}  \to \{x, \, \r\}$, 
as inspired by \cite{Ritus},  we end up with 
\bea
\G_{\rm I + II} &=& \frac{4e^2}{(4\p)^6 }
\int {\rm d}^8 z \, W^2 {\bar W}^2 
\int \limits_{0}^{\infty}  {\rm d}s
\int \limits_{0}^{\infty}  {\rm d}t 
\int \limits_{0}^{\infty} \frac{ {\rm d} u}{u^2} 
\sqrt{ \det
\left( \frac{s F}{\sinh(s F) }
\frac{t F}{\sinh(t F)} \right) } \non \\
& \times & 
\frac{\sin (sB/2)}{sB} \, \frac{\sin (s{\bar B} /2)}{s \bar B} \,
\frac{\sin (tB/2)}{tB} \, \frac{\sin (t{\bar B} /2)}{t \bar B}  \,
{\rm e}^{-{\rm i} ({\bar \F}\F -{\rm i}\ve) (s+t)}
\label{Gamma1}\\
& \times &
\int {\rm d}^4 \r \, 
\Big( \r \, \frac{F}{ \sinh( s F)}  \frac{F}{ \sinh( t F)} \,\r \Big)\,  
{\rm e}^{  {\rm i}  \r  A  \r /4 } ~, \non
\eea
where 
\be
A = F \coth (sF) + F \coth(tF) + {1 \over u}~.
\label{A}
\ee
The parallel displacement propagators that come from 
the two Green's functions in (\ref{final1}) annihilate each other, 
in accordance with (\ref{collapse}).

Using the explicit structure of the chiral kernel
(\ref{chiral-kernel}), it is easy to calculate 
the contribution from the third supergraph 
\bea
\G_{\rm III} &=& -\frac{8e^2}{(4\p)^6 }
\int {\rm d}^8 z \, W^2 {\bar W}^2 
\int \limits_{0}^{\infty} {\rm d}s
\int \limits_{0}^{\infty}  {\rm d}t 
\int \limits_{0}^{\infty} \frac{ {\rm d} u}{u^2} 
\sqrt{ \det
\left( \frac{s F}{\sinh(s F) }
\frac{t F}{\sinh(t F)} \right) } \non \\
& \times & 
\Big\{ \frac{\sin^2 (sB/2)}{(sB)^2} \,  
\frac{\sin^2 (t{\bar B} /2)}{(t{\bar B})^2}  ~+~ 
(s \leftrightarrow t) \Big\} \, {\bar \F} \F\,
{\rm e}^{-{\rm i} ({\bar \F}\F -{\rm i}\ve) (s+t)}
\label{Gamma2}\\
& \times &
\int {\rm d}^4 \r \,   
{\rm e}^{  {\rm i}  \r  A  \r /4 } ~. \non
\eea

${}$Following the non-supersymmetric consideration 
of Ritus \cite{Ritus}, 
it is useful to introduce the generating functional of 
Gaussian moments
\be
Z(p) = 
\frac{1}{(4\p)^2}
\int {\rm d}^4 \r \exp \left\{ 
\frac{\rm i}{4} \r_a \, A^a{}_b \,\r^b  + {\rm i} p_a \,\r^a \right\} 
= \frac{\rm i}{ \sqrt{\det A} } \, {\rm e}^{ -{\rm i} p \, A^{-1}\,  p }~,  
\ee
where $A$ is defined in (\ref{A}) and is
such that $\eta A = (\eta_{ab}\, A^b{}_c)$ is symmetric, 
with $\eta_{ab}$ the Minkowski metric. 
${}$From this we get  
two important special cases: 
\bea
\frac{1}{(4\p)^2} \int {\rm d}^4 \r \, 
{\rm e}^{  {\rm i}  \r  A  \r /4 } 
= Z(0)
&=&\phantom{-} \frac{\rm i}{ \sqrt{\det A} } ~,    \\
\frac{1}{(4\p)^2} \int {\rm d}^4 \r \, \r_a \r_b\, 
{\rm e}^{  {\rm i}  \r  A  \r /4 } 
= -\frac{\pa^2 }{\pa p^a \pa p^b} \,Z(p)\Big|_{p=0}
&=& - \frac{ 2 }{ \sqrt{\det A} } \, (A^{-1})_{ab}~. 
\eea
These results allow us to do the Gaussian $\r$-integrals
in (\ref{Gamma1}) and (\ref{Gamma2}).

As a next step, we have to compute the determinant of  $A$, 
with $A$ defined  in (\ref{A}),  as well as  the expression
$$
\tr \left[ 
\frac{F}{\sinh (sF)} \, \frac{F}{\sinh (tF)} \, A^{-1}\right]
$$
which appears in (\ref{Gamma1}) after having done the $\r$-integral.
Recalling the eigenvalues of $F=(F_a{}^b)$ given 
in eq. (\ref{eigenvalues}),  we  obtain 
\be
\frac{1}{ \sqrt{ \det A} } 
= \frac{1} {(u^{-1} +a_+)(u^{-1} +a_-)} ~,
\ee
where 
\be 
a_\pm = \l_\pm \coth(s\l_\pm) + \l_\pm \coth(t \l_\pm)~.
\label{a-pm}
\ee
With the notation 
\be
P_\pm = \frac{1}{st}\,
\frac{s \l_\pm} 
{\sinh (s\l_\pm)}\,\frac{t\l_\pm}{ \sinh(t \l_\pm)}~,
\label{P-pm}
\ee
we also get
\bea
&& \hf \,\tr \left[ 
\frac{F}{\sinh (sF)} \, \frac{F}{\sinh (tF)} \,
\frac{1}{F \coth (sF) + F \coth(tF) + u^{-1} } \right] \non \\
&& \qquad \qquad = \frac{P_+}{u^{-1} +a_+}
+ \frac{P_-}{u^{-1} +a_-} ~. 
\eea
With the old result (\ref{det1}), all the building blocks
in (\ref{Gamma1}) and (\ref{Gamma2}) thus become
simple functions of the $B$ and $\bar B$.

The proper-time $u$-integrals in 
(\ref{Gamma1}) and (\ref{Gamma2}) 
are identical to the ones considererd by Ritus
\cite{Ritus}.  Two integrals occur
\bea 
I_1(s,t)&=& \int\limits_{0}^{\infty} \frac{{\rm d}u } {u^2} \,
\frac{1} {(u^{-1} +a_+)(u^{-1} +a_-)}  ~, 
\label{I-1} \\
I_2(s,t)&=& \int\limits_{0}^{\infty} \frac{{\rm d}u } {u^2} \,
\frac{1} {(u^{-1} +a_+)(u^{-1} +a_-)} 
\left( \frac{P_+}{u^{-1} +a_+}
+ \frac{P_-}{u^{-1} +a_-} \right) ~,
\label{I-2}
\eea
and their direct evaluation gives
\bea
I_1(s,t)&=& \frac{1}{a_+ - a_- } \, 
\ln \Big( \frac{a_+}{a_-} \Big) ~, \\
I_2(s,t)&=&
\frac{1}{a_+ - a_- } \, 
\left( \frac{P_-}{a_-} - \frac{P_+}{a_+} \right) 
+  \frac{P_+ - P_-}{(a_+ - a_-)^2 } \, 
\ln  \Big( \frac{a_+}{a_-}\Big) ~.
\eea
However,  the expressions obtained do not make manifest
the fact that the two-loop effective action 
\bea
\G_{\rm two-loop} &=& \G_{\rm I+II} + \G_{\rm III}
=-\frac{e^2}{(4\p)^4 }
\int {\rm d}^8 z \, W^2 {\bar W}^2 
\int \limits_{0}^{\infty} {\rm d}s
\int \limits_{0}^{\infty}  {\rm d}t \,
{\rm e}^{-{\rm i} ({\bar \F}\F -{\rm i}\ve) (s+t)} \non \\
& \times &
\frac{ s\l_+}{\sinh (s\l_+)} \, \frac{ s\l_-}{\sinh (s\l_-)} \,
\frac{ t\l_+}{\sinh (t\l_+)} \, \frac{ t\l_-}{\sinh (t\l_-)} \non \\
&\times & \left(
\frac{\sin (sB/2)}{sB/2} \, \frac{\sin (s{\bar B} /2)}{s {\bar B}/2} \,
\frac{\sin (tB/2)}{tB/2} \, \frac{\sin (t{\bar B} /2)}{t {\bar B}/2}  \,
I_2(s,t) \right.
\label{two-loop-action} \\
&+& \left. ~\,
 \frac{\rm i}{2} {\bar \F}\F \, 
\Big\{ \frac{\sin^2 (sB/2)}{(sB/2)^2} \,  
\frac{\sin^2 (t{\bar B} /2)}{(t{\bar B}/2)^2}  ~+~ 
(s \leftrightarrow t) \Big\} \,I_1(s,t) \right) \non 
\eea
is free of any divergences, unlike the two-loop 
QED effective action \cite{Ritus}. 
This is why we would like to describe a different 
approach to computing the proper-time integrals, 
which is most efficient for evaluating effective actions  
in the framework of the derivative expansion.

The integrands in (\ref{I-1}) and (\ref{I-2}) involve 
two or three factors
of $(u^{-1} +a_\pm)^{-1}$, with $a_\pm$ defined in (\ref{a-pm}).
With the notation $x= st /u$, one can represent
\bea 
\frac{1} {(u^{-1} +a_\pm)} 
=  \frac{st }{x+s+t} \;
\left\{ 1 + 
\sum_{n=1}^{\infty} (-1)^n \,
\Big(\frac{st}{x+s+t}  \Big)^n
\, \Big( \cL_{\pm} (s) +  \cL_{\pm} (t) \Big)^n
\right\}~,
\eea
where 
\be
\cL_{\pm} (s) = 
\l_\pm \coth(s\l_\pm) -{1 \over s} 
\ee
is regular at $s=0$.  Using these decompositions and 
replacing the integration variable $u \to x=st/u$, 
one can easily do the integrals (\ref{I-1}) and (\ref{I-2}). 
Now, if one  takes into account the explicit form of $P_\pm$,  
see eq. (\ref{P-pm}), 
as well as the structure of the effective 
action (\ref{two-loop-action}), 
it is easy to see  that all the remaining proper-time 
$s$- and $t$-integrals 
are of the following generic form (after the Wick rotation 
$s = -{\rm i} \tilde{s} $ and  $t = -{\rm i} \tilde{t} $)
\bea
\int\limits_0^\infty {\rm d} \tilde{s} \int\limits_0^\infty {\rm d} \tilde{t}\,
\frac{\tilde{s}^m \tilde{t}^n }{(\tilde{s}+\tilde{t})^p} \, {\rm e} ^{-\m (\tilde{s}+\tilde{t})} 
=  \frac{ (m+n+1-p)! \,m! \,n!}{(m+n+1)!} \, {1 \over \m^{m+n +2 -p} }~, 
\qquad 
\m> 0~,
\label{funny-int}
\eea
with $m,n$ and $p$ non-negative integers such that $p \leq m+n+1$.

Recently, Dunne and Schubert \cite{DSch}
obtained closed-form expressions 
for the two-loop scalar and spinor QED effective 
Lagrangians in the case of a slowly varying 
{\it self-dual} background. 
In the supersymmetric case, the effective action 
vanishes for a self-dual vector multiplet. 
Nevertheless, the results of \cite{DSch} may be 
helpful in order to obtain a closed-form expression 
for a holomorphic part of  the two-loop effective action
(\ref{two-loop-action}) 
\bea
\G_{\rm holomorphic} &=& 
\int{\rm d}^8 z \,{ {\bar W}^2 W^2 \over  {\bar \F}^2\F^2 }\,  
\Big\{
 \L(\J^2) ~+~ {\bar \L}( {\bar \J}^2) \Big\} ~,
\eea
with $\J^2$ and ${\bar \J}^2$ defined in (\ref{psi}).

The effective action (\ref{two-loop-action}) contains 
supersymmetric extensions
of the terms $F^{2n}$, where $n=2,3,\ldots$, with $F$ 
the electromagnetic field
strength. Of special importance is the leading $F^4$ 
quantum correction, 
whose manifestly supersymmetric form is 
\be
c \int {\rm d}^8 z \, \frac{ {\bar W}^2  W^2}{{\bar \F}^2 \F^2}~.
\label{N=1F4}
\ee 
It can be singled out from (\ref{two-loop-action}) by 
considering the limit $B, \,{\bar B} \to 0$ in conjunction with 
$$
I_1(s,t) \to  \int\limits_{0}^{\infty} \frac{{\rm d}u } {u^2} \,
\frac{1} {(u^{-1} +s^{-1} + t^{-1})^2 } ~, \qquad 
I_2(s,t) \to {2\over st} \, \int\limits_{0}^{\infty} \frac{{\rm d}u } {u^2} \,
\frac{1} {(u^{-1} +s^{-1} + t^{-1})^3 } ~.
$$
Direct evaluation, with  use of (\ref{funny-int}), gives
\bea
c_{\rm two-loop}&=& \frac{e^2}{2(4\p)^4 }~.
\label{F4-two-loop}
\eea
This result turns out to be in conflict with a prediction 
made in \cite{BKO} on the basis of the background field 
formulation in $\cN=2$ harmonic superspace \cite{BBKO}.
According to \cite{BKO},  no $F^4$ quantum correction 
occurs at two loops in generic $\cN=2$ 
super Yang-Mills theories on the Coulomb branch.

Unfortunately,  the consideration of \cite{BKO} contains 
a subtle loophole.
Its origin will be uncovered in the next section. 
It will also be shown 
that a careful evaluation of two-loop 
$\cN=2$ harmonic supergraphs 
leads to the same result  (\ref{F4-two-loop}) 
we have just obtained from $\cN=1$ superfields.

\sect{The two-loop \mbox{$F^4$} quantum correction
 from harmonic supergraphs}

In this section, we will re-derive the two-loop $F^4$ 
quantum correction using an off-shell formulation for
$\cN=2$ SQED in harmonic superspace \cite{GIOS}.

The $\cN=2$ harmonic superspace ${\bR}^{4|8}\times S^2$
extends conventional superspace,
with coordinates $z^M = (x^m , \q^\a_i , {\bar \q}_\ad^i )$,
where $i =\underline{1},  \underline{2}$,
by the two-sphere $S^2 =SU(2)/U(1)$
parametrized by harmonics, i.e., group
elements
\bea
({u_i}^-\,,\,{u_i}^+) \in SU(2)~, \quad
u^+_i = \ve_{ij}u^{+j}~, \quad \overline{u^{+i}} = u^-_i~,
\quad u^{+i}u_i^- = 1 ~.
\eea
The main conceptual advantage of harmonic superspace
is that both the $\cN=2$ Yang-Mills vector multiplets and
hypermultiplets can be described by {\it unconstrained} 
superfields over the analytic
subspace of ${\bR}^{4|8}\times S^2$
 parametrized by the variables
$ \z^\cM \equiv (x^m_A,\q^{+\a},{\bar\q}^+_{\dot\a}, \,
u^+_i,u^-_j) $, 
where the so-called analytic basis is defined by
\be
x^m_A = x^m + 2{\rm i}\, \q^{(i}\s^m {\bar \q}^{j)}u^+_i u^-_j~, 
\qquad
 \q^\pm_\a=u^\pm_i \q^i_\a~, \qquad {\bar \q}^\pm_{\dot\a}
=u^\pm_i{\bar \q}^i_{\dot\a}~.
\ee
The $\cN=2$
Abelian vector multiplet is described
by a real analytic superfield $V^{++}  (\z)$.
The charged  hypermultiplet 
is described by an analytic superfield
$Q^+ (\z)$ and its conjugate $\breve{Q}^+ (\z)$.
The classical action for  $\cN=2$ SQED is 
\be
S_{\rm SQED }~ =~
\frac{1 }{2 e^2 }
 \int {\rm d}^4 x
{\rm d}^4 \q \,W^2
 - \int  {\rm d} \zeta^{(-4)}\,
\breve{Q}{}^+ \cD^{++}Q^+ ~.
\label{n=2sqed-action-har}
\ee
Here 
$W(z)$ is the $\cN=2$ 
chiral superfield strength \cite{GSW},
$ {\rm d} \zeta^{(-4)}$ denotes the analytic subspace
integration measure, 
and the harmonic (analyticity-preserving) 
covariant derivative is 
$\cD^{++}= D^{++} \pm  {\rm i} \,V^{++} $ when acting on 
$Q^+$ and $ \breve{Q}{}^+$, respectively.
The vector multiplet kinetic term in 
(\ref{n=2sqed-action-har}) can be expressed
as a gauge invariant functional of $V^{++}$ 
\cite{Zupnik}.

Upon quantization in the background field approach
\cite{BBKO},  the quantum theory is governed  
by the action  (lower-case letters are used for the 
quantum superfields)
\bea
S_{\rm quantum} &=&  \frac{1 }{2 e^2 }
\int  {\rm d} \zeta^{(-4)}\, v^{++} \Box \,v^{++} 
-   \int  {\rm d} \zeta^{(-4)}\,
\breve{ q}{}^+ \Big( \cD^{++} +{\rm i} \,v^{++}\Big) 
{q}^+ ~,
\eea
which has to be used for loop calculations.
The relevant Feynman propagators 
\cite{GIOS,BBKO} are
\bea
{\rm i} \,\langle 
{ q}^+(\z_1)\,\breve{ q}{}^+( \z_2) \rangle 
&=& \phantom{-}
 \frac{1}{{\stackrel{\frown}{\Box}}{}_1}\,
(\cD_1^+)^4 \,(\cD_2^+)^4\,
 \delta^{12}(z_1-z_2)
{1 \over (u^+_1 u^+_2)^3}~, \non \\
{\rm i} \,\langle 
v^{++}(\z_1)\,v^{++}( \z_2) \rangle &=& - {e^2 \over \Box_1}\,
\d_{\rm A}^{(2,2)} (\z_1,\z_2)
\label{harmonic-props}\\
&=& 
{ e^2  \over (4\p)^2} \int_0^\infty  \frac{{\rm d}s}{s^2} \,
(D_1^+)^4\,
{\rm e}^{ {\rm i}\r^2 /4s } \, 
\d^8(\q_1 -\q_2 ) \, \d^{(-2,2)}(u_1,u_2) ~,\non 
\eea
with $\d_{\rm A}^{(2,2)} (\z_1,\z_2)$
the analytic delta-function  \cite{GIOS}, 
\be
\d_{\rm A}^{(2,2)} (\z_1,\z_2) = 
(D_1^+)^4\,
\delta^{12}(z_1-z_2) \, \d^{(-2,2)}(u_1,u_2)~.
\label{ad-f}
\ee
Here the two-point function $\r^a$ is defined 
similarly to its $\cN=1$ counterpart  (\ref{two-point}).
The covariantly analytic d'Alembertian \cite{BBKO} is
\bea
{\stackrel{\frown}{\Box}}{}&=&
{\cal D}^m{\cal D}_m-
\frac{{\rm i}}{2}({\cal D}^{+\a}\cW){\cal D}^-_\a
- \frac{{\rm i}}{2}
({\bar{\cal D}}^+_{\dot\alpha}{\bar \cW}){\bar{\cal D}}^{-{\dot\alpha}}
+ \frac{{\rm i}}{4}({\cal D}^{+\a} {\cal D}^+_\a \cW) \cD^{--}\non \\
&{}& -\frac{{\rm i}}{8}[{\cal D}^{+\alpha},{\cal D}^-_\alpha] \cW
- {\bar \cW}\cW ~,
\eea
where $\cW = \pm W$ when acting on 
$q^+$ and $ \breve{q}{}^+$, respectively.
The algebra
of $\cN=2$ gauge covariant derivatives
$\cD_A = (\cD_a, \cD^i_\a , {\rm \cD}^\ad_j)
=D_A +{\rm i} \,\cA_A$
derived in  \cite{GSW} can be expressed
in the form
\bea
&
\{ \cD^+_\a ,  \cD^+_\b \}
= \{ {\bar \cD}^+_\ad , {\bar \cD}^+_\bd \}
=\{ \cD^+_\a , {\bar \cD}^+_\bd \} = 0~, \non \\
& \{ {\bar \cD}^+_\ad , \cD^-_\a \}
= - \{ \cD^+_\a , {\bar \cD}^-_\ad \}
=2{\rm i}\, \cD_{\a \ad}~,  \\
&\{ \cD^+_\a ,  \cD^-_\b \}
= -  2{\rm i}\,\ve_{\a \b} {\bar \cW}~, \qquad
\{ {\bar \cD}^+_\ad , {\bar \cD}^-_\bd \}
= 2{\rm i}\,\ve_{\ad \bd} \cW~, \non
\eea
where $\cD^\pm_\a = \cD^i_\a \,u^\pm_i$ and 
${\bar \cD}^\pm_\ad = {\bar \cD}^i_\ad \,u^\pm_i$.

Let us recall the argument given in \cite{BKO} that 
{\it no} non-holomorphic quantum corrections of the form 
\be
\int {\rm d}^{12}z  \,H(W,{\bar W}) \equiv
 \int {\rm d}^4 x
{\rm d}^8 \q \,H(W,{\bar W})
\label{non-hol}
\ee
occur at two loops. By definition, 
the two-loop effective action is
\bea
\G_{\rm two-loop} = { {\rm i}^3 \over 2} 
\int  {\rm d} \zeta^{(-4)}_1 
\int {\rm d} \zeta^{(-4)}_2 \,
\langle v^{++}(1)\,v^{++}( 2) \rangle \, 
\langle 
{ q}^+(1)\,\breve{ q}{}^+( 2) \rangle \,
\langle 
{ q}^+(2)\,\breve{ q}{}^+( 1) \rangle~,
\label{harm-2l-ea}
\eea
and it is generated by a single supergraph
 depicted in Figure 4.
\begin{figure}[!htb]
\begin{center}
\includegraphics{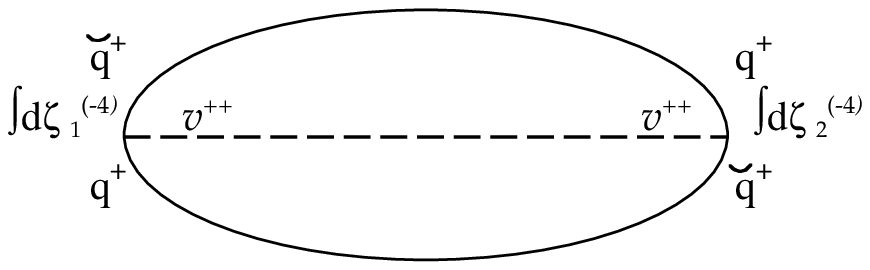}
\caption{Two-loop harmonic supergraph}
\end{center}
\end{figure}

${}$Following \cite{BKO},
the crucial step is to lift the analytic subspace 
integrals to those over the full superspace, 
by representing,  say, 
$\langle  { q}^+(1)\,\breve{ q}{}^+( 2) \rangle$ 
in the form 
\be
\langle  { q}^+(1)\,\breve{ q}{}^+( 2) \rangle
= (\cD_1^+)^4 \,(\cD_2^+)^4\, A^{(-3,-3)} (1,2)~, 
\qquad
A^{(-3,-3)} (1,2)  =
\frac{1}{{\stackrel{\frown}{\Box}}{}_1}\,
{ \delta^{12}(z_1-z_2)
 \over (u^+_1 u^+_2)^3}~, 
\label{A33}
\ee
and then using the standard identity 
\be
\int  {\rm d} \zeta^{(-4)}\, (D^+)^4 \, L (z,u) = 
\int {\rm d}^{12}z {\rm d} u \, L(z,u)~.
\ee
Since we are only after 
the quantum correction (\ref{non-hol}), 
it now suffices to approximate, in the 
resulting two-loop expression 
\be 
\int {\rm d}^{12}z_1 {\rm d} u_1
\int {\rm d}^{12}z_2 {\rm d} u_2\,
\langle v^{++}(1)\,v^{++}( 2) \rangle \, 
A^{(-3,-3)} (1,2)
\langle 
{ q}^+(2)\,\breve{ q}{}^+( 1) \rangle~,
\label{non-hol-2}
\ee
the covariantly analytic d'Alembertian 
by a  free massive one, 
\be
{\stackrel{\frown}{\Box}} \approx 
\Box -{\bar W}W~.
\label{approximation}
\ee
Now, the part of the integrand in (\ref{non-hol-2}), 
which involves the Grassmann delta-functions and 
spinor covariant derivatives, becomes
\be
\d^8(\q_1 - \q_2)\, \Big\{ (D^+_1)^4 \d^8(\q_1 - \q_2) \Big\} \,
 \Big\{ (D^+_1)^4 (D^+_2)^4 
\d^8(\q_1 - \q_2) \Big\} ~,
\ee
and this expression is obviously zero. Therefore, 
one naturally  concludes $H(W,{\bar W}) =0$.

Unfortunately, there is a subtle loophole in the above 
consideration. The point is that  
upon removing the two factors of $(D^+)^4$
from the hypermultiplet propagator  (\ref{A33}), 
in order to convert analytic integrals into 
full superspace integrals, we apparently end up with 
a more singular harmonic distribution, 
$A^{(-3,-3)} (1,2)$,  than the original propagator.
As a result, the expression (\ref{non-hol-2})
contains  the product of two harmonic distributions
\be
\d^{(-2,2)}(u_1,u_2) \; {1 \over (u^+_1 u^+_2)^3}~,
\ee
and such a  product is ill-defined. To make the consideration 
sensible, we have to regularize the harmonic distributions 
$\d^{(-2,2)} (u_1,u_2)$ and $(u^+_1u^+_2)^{-3}$ 
from the very beginning. However, the analytic delta-function 
(\ref{ad-f})
is known to be analytic in both arguments only 
if the right hand side involves the {\it genuine}
harmonic delta-function, see \cite{GIOS} for more details. 
With a regularized harmonic delta-function, however,
one has to use a modified (but equivalent) expression for 
the analytic delta-function  \cite{GIOS}
\bea 
\d_{\rm A}^{(2,2)} (\z_1,\z_2) &=& 
{1 \over 2 \Box_1} (D_1^+)^4\, 
(D_2^+)^4\, (D^{--}_2)^2\,
\delta^{12}(z_1-z_2) \, \d^{(-2,2)}(u_1,u_2)\non \\
&=& {1 \over 2 \Box_1} (D_1^+)^4\, (D_2^+)^4\, 
(D^{--}_1)^2\,
\delta^{12}(z_1-z_2) \, \d^{(2,-2)}(u_1,u_2)~. 
\label{harm-delta}
\eea
This expression is good in the sense that
it  allows for a regularized 
nonsingular harmonic delta-function. 
But it is more  singular in space-time than 
(\ref{ad-f}) -- an additional source for 
infrared problems in quantum theory, 
as will be demonstrated shortly.

Using the alternative representation  (\ref{harm-delta})
for the analytic delta-function, we would like to  undertake 
a second attempt to evaluate $H(W,{\bar W})$.
Let  us start again with the expression 
(\ref{harm-2l-ea})  for $\G_{\rm two-loop}$
in which the gluon propagator now reads
\be
{\rm i} \,\langle 
v^{++}(1)\,v^{++}( 2) \rangle = - {e^2 \over 2 (\Box_1)^2}\,
(D_1^+)^4\,  (D_2^+)^4\, 
(D^{--}_1)^2\,
\delta^{12}(z_1-z_2) \, \d^{(2,-2)}(u_1,u_2) ~.
\ee
In contrast to the previous consideration,
we now make use of the two factors of $(D^+)^4$ 
from  $ \langle v^{++}(1)\,v^{++}( 2) \rangle$ 
in order to convert the analytic subspace integrals into 
ones over the full superspace, thus leaving the hypermultiplet
propagators intact.  Such a procedure will lead, 
up to an overall numerical factor,  to 
\bea 
&& \int {\rm d}^{12} z_1 {\rm d} u_1
\int {\rm d}^{12}z_2 {\rm d} u_2\,
\Bigg\{  {1 \over ( \Box_1)^2} \, \d^{12}(z_1- z_2) \Bigg\} \,\non \\
& & \qquad \quad \times ~
\d^{(2,-2)} (u_1,u_2) \, 
(D^{--}_1)^2 \Bigg\{ 
\langle  { q}^+(1)\,\breve{ q}{}^+( 2) \rangle \,
\langle  { q}^+(2)\,\breve{ q}{}^+( 1) \rangle 
\Bigg\}~. 
\label{non-hol-3}
\eea
This does not seem to be identically zero and, 
in fact, can easily be evaluated.
The crucial step is to make use of the  identity \cite{KM2}
\bea
&& (\cD^+_1)^4 (\cD^+_2)^4 \;
\frac{\d^{12} (z_1 -z_2)  }{(u^+_1 u^+_2)^3}
\label{master}   \\
&& \qquad =
(\cD^+_1)^4 \;
\left\{ (\cD^-_1)^4 \,(u^+_1 u^+_2)
- \frac{\rm i}{2} \,
\D^{--}_1\;
(u^-_1 u^+_2)
- {\stackrel{\frown}{\Box}}_1 \;
\frac{(u^-_1 u^+_2)^2 }{(u^+_1 u^+_2)} \right\}
\;\d^{12} (z_1 -z_2)  ~,
\non
\eea
where
\bea
\D^{--} =\cD^{\a \ad} \cD^-_{\a} {\bar \cD}^-_{\ad}
&+& \hf \cW (\cD^-)^2 + \hf {\bar \cW} ({\bar \cD}^-)^2 \non \\
&+& (\cD^- \cW) \cD^- + ({\bar \cD}^- {\bar \cW}) {\bar \cD}^-
+\hf (\cD^- \cD^- \cW) ~.
\label{Delta--}
\eea
If we are only after $H(W,{\bar W})$, 
the covariantly analytic d'Alembertian can again 
be approximated as in  (\ref{approximation}).
Because of the Grassmann delta-function 
in the first line of  (\ref{non-hol-3}), 
only the first term in the right-hand side
of (\ref{master}) may produce a non-vanishing 
contribution. With the harmonic identities
\be
(u^+_1 u^+_2)|_{1=2} = 0~, \qquad
D^{--}_1 (u^+_1 u^+_2) = (u^-_1 u^+_2)~,
\qquad (u^-_1 u^+_2)|_{1=2} = -1~, 
\ee
it can be seen  that $H(W,{\bar W})$
is determined by the momentum integral 
\be 
H(W,{\bar W}) \propto
\int {\rm d}^4 p \int {\rm d}^4 k \; \frac{1}
{(p^2 + {\bar W}W) \, (k^2 + {\bar W}W)\, (p+k)^4}~.
\label{UV/IR}
\ee
The bad news is that this integral is both UV and IR 
divergent. This is the price one has to pay 
for having made use of the 
IR-unsafe representation (\ref{harm-delta}).  

It is of course possible to regularize the integral (\ref{UV/IR})
and, then, extract a finite part. Instead of practising 
black magic, however, we would like to present 
one more calculation that will lead to 
a manifestly finite and well-defined 
expression for $H(W,{\bar W}) $. 
The idea is to take seriously 
the representation (\ref{harm-2l-ea}) 
and stay in the analytic subspace at all
stages of the calculation, without 
 artificial conversion of analytic  integrals
into those over the full superspace
(and without  use of the 
IR-unsafe representation (\ref{harm-delta})).
Instead of computing the contribution 
(\ref{non-hol}) directly,  in such a setup 
we should actually look for 
an equivalent higher-derivative quantum correction
of the form
\be
\int {\rm d}\z^{(-4)}  \,(D^+)^4 H(W,{\bar W})~.
\label{non-hol-anal}
\ee
We are going to work with 
 an on-shell $\cN=2$ vector multiplet 
background
\be
D^+ D^+\, W = 0~.
\ee

In the analytic basis,  the  delta-function 
(\ref{ad-f}) can be represented as \cite{GIOS}
\be
\d_{\rm A}^{(2,2)} (\z_1,\z_2) 
= \d^4 (x_1-x_2) \, 
\Big(\q^+_1 - (u^+_1 u^-_2) \q^+_2 \Big)^4 \,
\d^{(-2,2)}(u_1,u_2) ~.
\ee
Let us  use this expression 
for   $\d_{\rm A}^{(2,2)} (\z_1,\z_2) $ 
in the gluon propagator
$\langle v^{++}(\z_1)\,v^{++}( \z_2) \rangle$, 
as defined in eq.  (\ref{harmonic-props}), 
which appears in the effective action  (\ref{harm-2l-ea}).
It is obvious that the operator $(1 /\Box_1)$ 
acts  on $\d^4 (x_1-x_2)$ only.
The Grassmann delta-function,
$ \Big(\q^+_1 - (u^+_1 u^-_2) \q^+_2 \Big)^4 $, 
can be used to do one of the Grassmann 
integrals in  (\ref{harm-2l-ea}). 
Similarly, the harmonic delta-function, 
$\d^{(-2,2)}(u_1,u_2) $, can be used to do 
one of the harmonic integrals 
in (\ref{harm-2l-ea}). 
As a result, the hypermultiplet propagators in 
(\ref{harm-2l-ea}) should be evaluated 
in the following coincidence limit: 
$\q_1 = \q_2 $ and $u_1 = u_2$.
To implement this limit,  it is again advantageous
to make use of the identity (\ref{master}).
It is not difficult to see that only the second 
term on the right of (\ref{master}) can contribute.
Each term in the operator $\D^{--}$, (\ref{Delta--}), 
contains two spinor derivatives. 
Taken together with the overall factor
$(\cD^+)^4$ in (\ref{master}), we 
have a total of six spinor derivatives. 
But we need eight such derivatives to annihilate 
the spinor delta-function $\d^8(\q_1 -\q_2)$
entering each hypermultiplet propagator.
Two missing derivatives come from 
the covariantly analytic d'Alembertian. Introducing 
the Fock-Schwinger proper-time representation
\be
-\frac{1}{ {\stackrel{\frown}{\Box}}}= {\rm i}
\int_{0}^{\infty}{\rm d}s\;
{\rm e}^{{\rm i}\,s\,{\stackrel{\frown}{\Box}}}~,
\ee
it turns out to be  sufficient to  approximate 
\bea
{\rm e}^{{\rm i}\,s{\stackrel{\frown}{\Box}}} 
& \approx & 
\hf \Big({s\over 2}\Big)^2 \,
\Big\{ 
({D}^{+\a}W){\cal D}^-_\a
+ ({\bar{D}}^+_{\dot\alpha}{\bar W}){\bar \cD}^{-{\dot\alpha}}
\Big\}^2 \, 
{\rm e}^{{\rm i}\,s(\Box- {\bar W}W)}~.
\eea
After that, it only remains to apply the identity
\be
(D^+)^4\,(D^-)^4\, \d^8(\q-\q')\Big|_{\q=\q'}=1
\ee
in order to  complete the $D$-algebra gymnastics.
The remaining technical steps (i.e. the calculation of Gaussian 
space-time integrals and of triple proper-time
integrals) are identical to those described before 
in the $\cN=1$  case. Therefore, we simply 
give the final result
for the quantum correction under consideration:
\bea
\frac{e^2}{32(4\p)^4}
\int  {\rm d} \zeta^{(-4)}\,
\frac{(D^+ W)^2 ({\bar D}^+ {\bar W})^2 }
{(W {\bar W})^2} 
&=& \frac{e^2}{2(4\p)^4}
\int  {\rm d} \zeta^{(-4)}\,
(D^+)^4 \Big( \ln W \,\ln {\bar W} \Big) \non \\
&=&  \frac{e^2}{2(4\p)^4}
\int {\rm d}^4 x {\rm d}^8 \q \,
\ln W \,\ln {\bar W} ~.
\eea
Upon reduction to $\cN=1$ superspace, 
this functional can be shown to take the form 
(\ref{N=1F4}) with the coefficient $c$ equal to 
(\ref{F4-two-loop}). 
The reduction to $\cN=1$ superfields
is defined as usual:
$U| = (x,\q_i, {\bar \q}^j )|_{
\q_{\underline{2}}
= {\bar \q}^{\underline{2}} =0}$, 
for any $\cN=2$ superfield $U$.
The $\cN=1$ components of $W$ are 
\be
\F = W|~, \qquad  
-2{\rm i}\, W_\a = D_\a^{\underline{2}} \,W|~.
\ee

Harmonic superspace still remains to be 
tamed for quantum practitioners, 
and the present situation is reminiscent  of  
that with   QED in the mid 1940's.  It is worth hoping that, 
as with  QED, it should take no longer than half a decade
of development for this approach to become 
a safe and  indispensable 
scheme for  quantum calculations in $\cN=2$ SYM theories.

\vskip.5cm

\noindent
{\bf Acknowledgements:}\\
We thank Jim Gates for  discussions and  
Arkady Tseytlin for helpful comments and suggestions.
S.M.K. is grateful to  the Department of Physics 
at the Ohio State University, 
C.N. Yang Institute for Theoretical Physics and 
the Center for String and Particle Theory 
at the University of Maryland
for hospitality during the course of the project.  
This work is supported in part by the Australian Research
Council, the Australian Academy of Science 
as well as by UWA research grants.

\begin{appendix}

\sect{Parallel displacement propagator}
In this appendix we describe, following \cite{KM}, 
the salient properties of the $\cN=1$ 
parallel displacement propagator
$I(z,z')$. This object is uniquely specified by 
the following requirements:\\
(i) the gauge transformation law
\be
 I (z, z') ~\to ~
{\rm e}^{{\rm i} \t(z)} \,  I (z, z') \,
{\rm e}^{-{\rm i} \t(z')} ~
\label{super-PDO1}
\ee
with respect to  an  arbitrary gauge ($\t$-frame) transformation 
of  the covariant derivatives
\be
\cD_A ~\to ~ {\rm e}^{{\rm i} \t(z)} \, \cD_A\,
{\rm e}^{-{\rm i} \t(z)}~, \qquad 
\t^\dagger = \t ~, 
\label{tau}
\ee
with the gauge parameter $\t(z)$ being arbitrary modulo 
the reality condition imposed;\\
(ii) the equation  
\be
\z^A \cD_A \, I(z,z') 
= \z^A \Big( D_A +{\rm i} \, \cA_A(z) \Big) I(z,z') =0~;
\label{super-PDO2}
\ee
(iii) the boundary condition 
\be 
I(z,z) ={\bf 1}~.
\label{super-PDO3}
\ee
These imply the important relation
\be
I(z,z') \, I(z', z) = {\bf 1}~,
\label{collapse}
\ee
as well as 
\be
\z^A \cD'_A \, I(z,z') 
= \z^A  \Big( D'_A \,I(z,z') 
 - {\rm i} \, I(z,z') \, \cA_A(z') \Big) =0~.
\ee
Under Hermitian conjugation, the parallel 
displacement propagator transforms as 
\be
\Big( I(z,z')\Big)^\dagger = I(z',z)~.
\ee

${}$For a covariantly constant vector multiplet, 
\be
\cD_a \, W_\b =0~, 
\label{constant SYM}
\ee 
the covariant differentiation of  $\cD_A \,I(z,z')$ gives \cite{KM}
\bea
\cD_{\b \bd} I(z,z') &=& I(z,z') \,
\Big( -  \frac{{\rm i}}{4} \r^{\ad \a} \cF_{\a \ad, \b \bd}(z') 
-{\rm i} \, \z_{\b} \bar{\cW}_{\bd}(z') 
+ {\rm i} \, \bar{\z}_{\bd} \cW_{\b}(z')
\nonumber \\ 
&& \phantom{I(z,z') \,\Big( }
+\frac{2{\rm i}}{3} \, 
\bar{\z}_{\bd}
\z^{\a}
\cD_{\a}\cW_{\b}(z')
+ \frac{2{\rm i}}{3} \, \z_{\b} \bar{\z}^{\ad}
\bar{\cD}_{\ad} \bar{W}_{\bd}(z') \Big) \nonumber \\
&=& \Big( - \frac{{\rm i}}{4} \r^{\ad \a} \cF_{\a \ad, \b \bd}(z) 
-{\rm i} \, \z_{\b} \bar{\cW}_{\bd}(z) 
+ {\rm i} \, \bar{\z}_{\bd} 
\cW_{\b}(z)
\nonumber \\ 
&  & 
\phantom{\Big(}
-\frac{{\rm i}}{3} \, \bar{\z}_{\bd} \z^{\a} \cD_{\a}\cW_{\b}(z) 
- \frac{{\rm i}}{3} \, \z_{\b} \bar{\z}^{\ad}
\bar{\cD}_{\ad} \bar{\cW}_{\bd}(z) \Big) \, I(z,z')~;
\label{super-PTO-der2} \\
\cD_{\b} I(z,z') &=& I(z,z') \,
\Big( \frac{1}{12} \, \bar{\z}^{\bd} 
\r^{\a \ad}  \cF_{\a \ad, \b \bd}(z') 
- {\rm i} \, \r_{\b \bd} \Big\{
\frac{1}{2} 
\bar{\cW}^{\bd}(z') 
-\frac{1}{3}  
\bar{\z}^{\ad} \bar{\cD}_{\ad}
\bar{\cW}^{\bd}(z')  \Big\}
\nonumber \\ 
&  & 
+\frac13 \, \z_{\b} \bar{\z}_{\bd} \bar{W}^{\bd}(z')  
 +\frac13
\bar{\z}^2 \Big\{ 
\cW_{\b}(z') + \hf \z^{\a} \cD_{\a}
\cW_{\b}(z')
-\frac{1}{4}  \z_{\b} \cD^{\a} \cW_{\a}(z') \Big\}
\Big)
\nonumber \\
&=& \Big( \frac{1}{12} \, \bar{\z}^{\bd} 
\r^{\a \ad} \, \cF_{\a \ad , \b \bd}(z)
- \frac{{\rm i}}{2} \, \r_{\b \bd} 
\Big\{ 
\bar{\cW}^{\bd}(z) 
+\frac{1}{3}  \bar{\z}^{\ad} \bar{\cD}_{\ad}
\bar{\cW}^{\bd}(z) \Big\} 
+ \frac13 \, \z_{\b} \bar{\z}_{\bd} \bar{\cW}^{\bd}(z)  \nonumber \\ 
& & \phantom{ \Big( } +\frac13
\bar{\z}^2 \Big\{
\cW_{\b}(z) - \hf  \z^{\a} \cD_{\a} \cW_{\b}(z)
+\frac{1}{4}  \z_{\b} {\cD}^{\a} {\cW}_{\a}(z) \Big\}
\Big)\, I(z,z')~;
\label{super-PTO-der3} \\
{\bar \cD}_{\bd} I(z,z') &=& I(z,z') \,\Big( 
-\frac{1}{12} \, \z^{\b } 
\r^{\a \ad} \, F_{\a \ad , \b \bd}(z')
- {\rm i} \,\r_{\b \bd} \Big\{ 
\frac{1}{2}  \cW^{\b}(z') 
+\frac{1}{3}  \z^{\a} \cD_{\a} \cW^{\b}(z')  \Big\}
\nonumber \\ &  & 
- \frac13 \, \bar{\z}_{\bd} \z^{\b} \cW_{\b}(z')  
-\frac13 \z^2 \Big\{ 
\bar{\cW}_{\bd}(z')  -\hf \bar{\z}^{\ad} \bar{\cD}_{\ad}
\bar{\cW}_{\bd}(z')
+\frac{1}{4}  \bar{\z}_{\bd} {\bar \cD}^{\ad} 
{\bar \cW}_{\ad}(z') \Big\} 
\Big)
\nonumber \\
&=& \Big( - \frac{1}{12} \, \z^{\b} 
\r^{\a \ad}  \cF_{\a \ad , \b \bd}(z)
- \frac{{\rm i}}{2} \, \r_{\b \bd} \Big\{ \cW^{\b}(z) 
- \frac{1}{3}  \z^{\a} \cD_{\a} \cW^{\b}(z)  \Big\} 
- \frac13 \, \bar{\z}_{\bd} \z^{\b} \cW_{\b}(z)  \nonumber \\ 
& & - \frac13 \z^2 \Big\{ 
\bar{\cW}_{\bd}(z) +\hf  \bar{\z}^{\ad} \bar{\cD}_{\ad}
\bar{\cW}_{\bd}(z)
-\frac{1}{4}  \bar{\z}_{\bd} {\bar \cD}^{\ad} 
{\bar \cW}_{\ad}(z) \Big\}
\Big) \, I(z,z')~.
\label{super-PTO-der4} 
\eea

\end{appendix}

\end{document}